# Student Satisfaction Mining in a Typical Core Course of Computer Science

Farzana Afrin, Mohammad Saiedur Rahaman, Mohammad Saidur Rahman and Mashiour Rahman

*Abstract*—Student's satisfaction plays a vital role in success of an educational institute. Hence, many educational institutes continuously improve their service to produce a supportive learning environment to satisfy the student need. For this reason, Educational institutions collect student satisfaction data to make decision about institutional quality, but till now it cannot be determined because student satisfaction is a complex matter which is influenced by variety of characteristics of students and institutions. There are many studies have been performed to inspect student satisfaction in the form of college services, programs, student's accommodation, student–faculty interaction, consulting hours etc. So, still we cannot have a standard method to know what is going on about satisfaction in the case of a core course. In this research we determined the attributes that heavily affect student satisfaction in a core course of computer science and the current status of other attributes as well.

## 1. Introduction

From the past two decade attention has been paid to research education by institutions of higher education, governments, and industry and professional partners, both nationally and internationally. Basically, Students are main concern who paid their attention and potential to achieve the knowledge and education from those institutions.

Student usually looks for high-quality institution where they can reflect their ability and performance and the parents who pay for also concern about it. In evaluating the effectiveness of continuing professional education programs, it is necessary to consider both program and work outcomes.

**Farzana Afrin** is a student of Master of Computer Science, Department of Computer Science, American International University-Bangladesh, Kamal Ataturk Avenue, Dhaka, Bangladesh. Email: farzana@aiub.edu
**Mohammad Saiedur Rahaman** is an Assistant Professor of Department of Computer Science, American International University-Bangladesh, Kamal Ataturk Avenue, Dhaka, Bangladesh. Email: saied@aiub.edu
**Mohammad Saidur Rahman** is an Assistant Professor of Department of Computer Science, American International University-Bangladesh, Kamal Ataturk Avenue, Dhaka, Bangladesh. Email: saidur@aiub.edu
**Mashiour Rahman** is an Assistant Professor of Department of Computer Science, American International University-Bangladesh, Kamal Ataturk Avenue, Dhaka, Bangladesh. Email: mashiour@aiub.edu

In [1], authors distinguish between the concept of teaching evaluation and teaching effectiveness. In their view, teaching evaluation focuses on whether teaching leads to the desired level of expertise in learning as indicated by the teaching objectives.

Teaching evaluation is concerned with issues of measurement and design, the accomplishment of learning objectives, and the attainment of requisite knowledge and skills which in fact reflects student's perceptions on the quality services received. In contrast teaching effectiveness deals with broader issues involved in understanding why the style of teaching did or did not lead to the desired level of proficiency in the practical field [1].

In [2], authors set an evaluation mode of teachers' teaching quality from aspects of different evaluation methods: Reflective teaching evaluation, professional group discussion, teaching archive etc. [3] attempts to provide basis and foundation for the construction of IT-based teaching evaluation system.

In [4], authors compared two universities, and revealed the difference of the course design concept, the indicators of teaching methods and the purpose of teaching evaluation system.

But an outcome of student learning also depends on the satisfaction of a student. Still it is undefined that what actual student satisfaction is. This research consults with the students and experienced teachers about it and come up with the answer about the factors that influence student's satisfaction in learning a core course of computer science.

The paper is organized as follows. Section 2 presents the proposed method. Some analytical results and recommendations are presented in Section 3. Finally, Section 4 concludes the paper. Also an appendix with sample questionnaire and dataset is attached at the end.

## 2. Proposed Method

### 2.1. Research Approach

To do the research, we collected information from the honor's level student (CS, COE, SE, CSSE, CSE, and CIS) of AIUB. All participation was voluntary. If the participants wished to withdraw, they were free to do so without providing any further information. We provided a message to all participants explaining the context of the study. As a candidate of core course of computer science we selected 'Algorithm' course as every student of the above mentioned disciplines have to complete this course.

### 2.2. Sampling Method

A random sampling method was adopted to collect data from the respondents. The random sampling was accomplished by selecting elements from a randomly arranged sample frame according to ordered criteria [5]. In [6], author mentioned that this particular process of choosing sample from the overall population saved time as well as financial and human resources. We surveyed on the students study at AIUB for the current study.

The study was limited only in AIUB because of time and economical restrictions. The research only considered Honor's level student. For this we used random sampling method. In this method of sampling, every element has a known and has equal chance of being selected.

For questionnaire survey, we selected every 40 students from the students' list provided by AIUB. For example, the sample includes the 40 respondent, the 80 respondent, the 120 respondents and so forth. A total of 500 questionnaires were distributed through online and 429 completed (usable) questionnaires were considered for statistical analyses and research hypotheses testing.

### 2.3. Research Instruments

To obtain information from the student of AIUB we used the online questionnaire method. The online questionnaire method has been considered as an appropriate and convenient method to collect data from student for the following reasons. First the current study was about practical ness, up-to-date-ness, disciplinarian, dynamism and enthusiasm of faculty and student might feel uncomfortable to discuss these issues with an interviewer. A questionnaire survey can be considered as a better option in such a condition as it ensures anonymity [7]. Second, we have conducted a questionnaire survey on 500 respondents. To interview such a large number of respondents would have been expensive as well as time consuming. For this reason, we conducted questionnaire survey rather than direct interview. Third, hard copy of questionnaire cost is very expensive. Forth, mail survey would have been too much time consuming. As a result, online survey was feasible for the current study.

### 2.4. Survey Instruments

In this research, the researcher used a structured questionnaire to collect information from the students of AIUB. The first section of the questionnaire focused on the practical and up-to-date-ness, the second section on the teacher disciplinarian and the third section on the dynamism and enthusiasm of faculty.

### 2.5. Data Collection Procedure

After obtaining the approval from AIUB, we launched this questionnaire on the web for the respondents. The respondents for this study were drawn from the student list, provided by AIUB lab authority. A total of 500 forms are filled up by student and 429 completed (usable) questionnaires were considered for statistical analyses and research hypotheses testing.

We uploaded the survey questionnaires among the respondents. This approach was adopted for a number of reasons. First number of internet users is growing at AIUB. As a result it was possible for the research to conduct internet survey. Second the research conducted a survey on 500 respondents and to interview such a large sample of respondents would have been time consuming and difficult. Time constraint was another factor for this reason here we used this approach of data collection for the study.

### 2.6. Data Analysis Procedures

To examine the total dataset, we used statistical techniques like correlation analysis [8][9]. Correlation analysis was employed to determine whether relationships exist between measured variables or not. Besides that using mean technique [10][11], exploratory analysis can be performed particularly for handling large scale of data for finding weights of attributes.

### 2.7. Proposed method

Figure 1 illustrates the work flow of proposed method. Initially the data collected from the answers of survey questionnaire is stored in a database. Next, from the database, a student opinion dataset is found. From this dataset correlation is calculated with students' satisfaction and other attributes. The attributes with higher positive correlation is most influential attributes.

The student opinion dataset is also sent for calculating the mean to know the weight of each attributes. This gives the current status of different attributes.

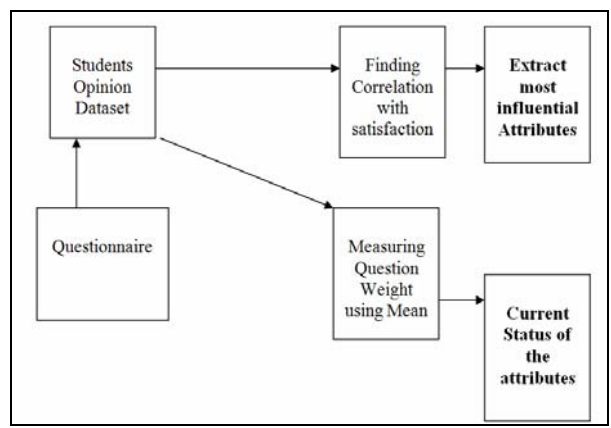

**Figure 1: Work flow of the proposed method**

## 3. Results and Discussion & Recommendations

This section covers the different type of analyses and findings of the study. Correlation and Mean analyses were employed to examine research outcome. The results of correlation analysis provided full support to all research outcomes. The results of Mean analysis however, provided full support to this research and bring desire expected result.

From table 1, we can see that satisfaction heavily influenced by well planning and organization, satisfactory course coverage, encouragement to participate in the class, and relation with real world software environment as they have maximum positive correlation with satisfaction

**Table 1: Correlation value with satisfaction and other attributes**

| Attributes | Correlation with overall satisfaction |
|---|---|
| Contains Intellectual grasp: | 0.44034571 |
| Appropriateness for CS program | 0.36229701 |
| Has Relation with real life | 0.49602112 |
| Planning and organization is well | 0.74037554 |
| Coverage is satisfactory | 0.67720576 |
| Counseling hours are Satisfactory | 0.51415581 |
| Encouragement in class participation | 0.63945366 |
| Sufficient number of cases | 0.42861021 |
| Relation with real world SD | 0.64279277 |
| Support from Prerequisite course | 0.46476726 |

Figure 2 says that students think that the course is appropriate for computer science program. But except this one other attributes has very low average and we can say that current situation is not satisfactory for all other attributes.

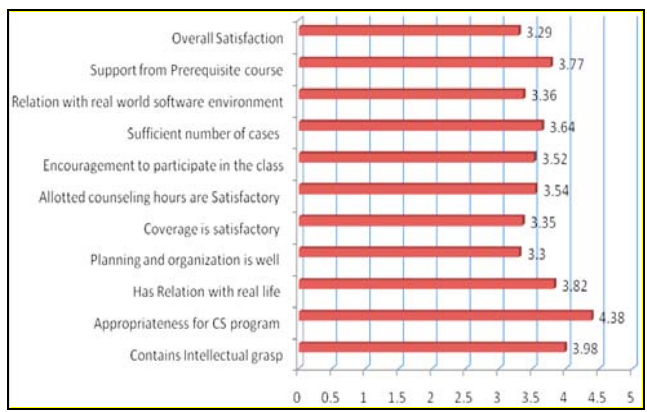

**Figure 2: Mean value of different attributes**

Educational organization should implement the findings of the research for the betterment of student. Well planning and organization, and satisfactory course coverage, can increase students satisfaction over a course. Moreover if a teacher can encourage a student to participate in the class, and relate the course with real world software environment then it will be an additional input to the student satisfaction. Along with these, we need to take care of other attribute that influence student satisfaction.

## 4. Conclusions

A method for finding the best attributes in students' satisfaction in core course of computer science has been proposed in this research. However, this research can be applied in other courses also. If data from several universities can be collected together, then more accurate result can be found. Though, we considered only the attributes that influences in how to satisfy a student in teaching a core course, other attributes like exam scripts evaluation, question setting standard can also be added for getting a better picture of the scenario.

**Appendix-A: Sample Questionaire and Part of dataset**

Following table shows the sample questionnaire. Here each carries points 1 through 5 where 1= Strongly disagree, 2=Disagree, 3=Neutral 4=Agree, 5=Strongly Agree. The table also shows the meaning of each question in the dataset.

| Attributes | Appeared in the dataset as |
|---|---|
| Contains Intellectual grasp: | Q1 |
| Appropriateness for CS program | Q2 |
| Has Relation with real life | Q3 |
| Planning and organization is well | Q4 |
| Coverage is satisfactory | Q5 |
| Counseling hours are Satisfactory | Q6 |
| Encouragement in class participation | Q7 |
| Sufficient number of cases | Q8 |
| Relation with real world SD | Q9 |
| Support from Prerequisite course | Q10 |
| Overall satisfaction | Q11 |

Following table exhibits some part of obtained dataset from the survey questionnaire with additional column 'GRADE' and corresponding GPA for those grades. This is used to check the reliability of the dataset.

| Q1 | Q2 | Q3 | Q4 | Q5 | Q6 | Q7 | Q8 | Q9 | Q10 | Q11 | GRADE | GPA |
|---|---|---|---|---|---|---|---|---|---|---|---|---|
| 5 | 4 | 5 | 3 | 3 | 4 | 4 | 4 | 5 | 4 | 4 | A | 3.75 |
| 5 | 5 | 3 | 3 | 2 | 2 | 3 | 3 | 2 | 2 | 4 | A | 3.75 |
| 4 | 3 | 2 | 3 | 3 | 2 | 4 | 3 | 2 | 2 | 4 | A | 3.75 |
| 4 | 5 | 5 | 4 | 3 | 3 | 1 | 1 | 4 | 2 | 2 | A | 3.75 |
| 4 | 5 | 4 | 3 | 3 | 4 | 4 | 4 | 3 | 3 | 3 | A | 3.75 |
| 4 | 4 | 4 | 4 | 4 | 4 | 4 | 4 | 4 | 4 | 4 | A- | 3.5 |
| 4 | 5 | 4 | 3 | 4 | 4 | 2 | 4 | 5 | 5 | 4 | A- | 3.5 |
| 5 | 5 | 4 | 4 | 4 | 3 | 4 | 3 | 4 | 4 | 3 | A- | 3.5 |
| 3 | 5 | 5 | 2 | 1 | 3 | 2 | 4 | 1 | 1 | 1 | A- | 3.5 |
| 4 | 5 | 4 | 4 | 2 | 4 | 4 | 4 | 4 | 4 | 3 | A- | 3.5 |
| 4 | 5 | 5 | 4 | 5 | 4 | 5 | 3 | 4 | 5 | 4 | A- | 3.5 |
| 5 | 5 | 4 | 5 | 4 | 5 | 5 | 5 | 4 | 5 | 5 | A+ | 4 |
| 4 | 4 | 3 | 3 | 3 | 4 | 3 | 4 | 3 | 3 | 2 | B | 3 |
| 5 | 5 | 2 | 1 | 1 | 1 | 1 | 2 | 4 | 1 | 3 | B | 3 |
| 4 | 3 | 1 | 2 | 2 | 5 | 3 | 2 | 2 | 2 | 2 | B | 3 |
| 4 | 5 | 4 | 4 | 4 | 4 | 4 | 3 | 4 | 4 | 4 | B | 3 |
| 3 | 3 | 3 | 4 | 4 | 3 | 4 | 4 | 4 | 4 | 4 | B | 3 |
| 5 | 5 | 5 | 5 | 5 | 5 | 5 | 4 | 5 | 5 | 5 | B | 3 |
| 5 | 5 | 2 | 4 | 4 | 3 | 5 | 4 | 5 | 3 | 4 | B | 3 |
| 4 | 4 | 4 | 3 | 2 | 3 | 3 | 4 | 3 | 4 | 4 | B | 3 |
| 5 | 5 | 5 | 3 | 4 | 1 | 5 | 3 | 5 | 5 | 5 | B | 3 |
| 4 | 5 | 4 | 4 | 5 | 5 | 5 | 4 | 4 | 4 | 4 | B | 3 |
| 4 | 4 | 4 | 4 | 2 | 5 | 5 | 4 | 4 | 4 | 4 | B | 3 |
| 5 | 4 | 4 | 5 | 5 | 4 | 4 | 5 | 4 | 4 | 5 | B | 3 |
| 4 | 5 | 5 | 4 | 3 | 3 | 3 | 4 | 4 | 4 | 5 | B- | 2.75 |
| 3 | 4 | 5 | 4 | 2 | 4 | 4 | 4 | 2 | 3 | 4 | B- | 2.75 |
| 1 | 1 | 1 | 1 | 1 | 3 | 1 | 1 | 1 | 1 | 1 | B- | 2.75 |
| 4 | 5 | 2 | 3 | 4 | 4 | 4 | 5 | 2 | 3 | 4 | B- | 2.75 |
| 4 | 4 | 5 | 3 | 3 | 4 | 3 | 3 | 3 | 2 | 4 | B- | 2.75 |
| 4 | 3 | 5 | 3 | 4 | 3 | 3 | 3 | 4 | 3 | 3 | B- | 2.75 |
| 4 | 3 | 4 | 3 | 2 | 3 | 2 | 4 | 4 | 4 | 3 | B- | 2.75 |
| 3 | 4 | 1 | 1 | 2 | 4 | 2 | 3 | 3 | 1 | 2 | B+ | 3.25 |
| 4 | 5 | 4 | 4 | 4 | 4 | 5 | 4 | 4 | 4 | 4 | B+ | 3.25 |
| 4 | 4 | 1 | 3 | 4 | 4 | 4 | 2 | 1 | 2 | 5 | B+ | 3.25 |
| 2 | 4 | 2 | 1 | 1 | 1 | 1 | 4 | 1 | 1 | 4 | B+ | 3.25 |
| 5 | 5 | 5 | 5 | 5 | 5 | 5 | 5 | 5 | 5 | 5 | B+ | 3.25 |
| 5 | 5 | 4 | 2 | 3 | 3 | 4 | 2 | 2 | 2 | 5 | B+ | 3.25 |
| 5 | 4 | 4 | 3 | 1 | 2 | 5 | 5 | 4 | 2 | 2 | B+ | 3.25 |
| 4 | 4 | 4 | 4 | 4 | 4 | 4 | 4 | 4 | 4 | 4 | B+ | 3.25 |
| 3 | 5 | 5 | 3 | 2 | 4 | 2 | 4 | 2 | 4 | 3 | B+ | 3.25 |
| 4 | 4 | 4 | 3 | 3 | 3 | 3 | 3 | 3 | 3 | 3 | C | 2 |
| 4 | 4 | 4 | 3 | 4 | 4 | 4 | 3 | 4 | 4 | 4 | C | 2 |
| 4 | 4 | 4 | 4 | 4 | 4 | 4 | 4 | 4 | 4 | 4 | C | 2 |
| 2 | 3 | 2 | 1 | 3 | 4 | 2 | 4 | 2 | 1 | 3 | C | 2 |
| 4 | 5 | 5 | 4 | 3 | 5 | 5 | 3 | 4 | 4 | 4 | C+ | 2.25 |
| 4 | 5 | 4 | 3 | 4 | 3 | 5 | 4 | 5 | 4 | 4 | C+ | 2.25 |
| 5 | 5 | 5 | 5 | 5 | 4 | 3 | 3 | 5 | 5 | 4 | C+ | 2.25 |

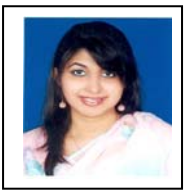
**Farzana Afrin** earned his B.Sc. in Computer Engineering in 2010 and M.Sc. in Computer Science in 2012 from the American International University – Bangladesh (AIUB). Currently she is with the Department of VUES, AIUB. Her research interests include data mining, and design & analysis of information systems.

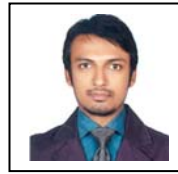
**Mohammad Saiedur Rahaman** earned his B.Sc. in Computer Science & Engineering in 2007 and M.Sc. in Computer Science in 2009 from the American International University – Bangladesh (AIUB). Currently he is with the Department of Computer Science, AIUB as an assistant professor. He authored several research papers in reputed journals. His research interests include wireless sensor networks, grid computing, mobile & multimedia communication and data mining.

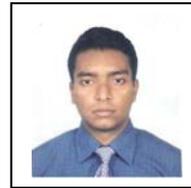
**Mohammad Saidur Rahman** completed his B.Sc. in Computer Engineering in 2007 and M.Sc. in Computer Science in 2009 from the American International University – Bangladesh (AIUB). Currently he is with the Department of Computer Science, AIUB as an assistant professor. He authored several research papers in reputed journals. His research interests include wireless mesh networks, wireless sensor networks, and mobile & multimedia communication.

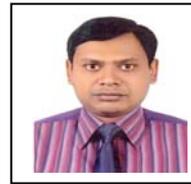
**Mashiour Rahman** completed his B.Sc. in Computer Science in 2000 from American International University-Bangladesh and Masters in Computing in 2005 from National University of Singapore. Currently he is Assistant Professor and Head of the Undergraduate program of Department of Computer Science at American International University-Bangladesh. He has authored several research papers in reputed journals. His research interests include Algorithm, Image Processing, Software Engineering, Distributed Computing and communication.